# Model of deep non-volcanic tremor part II: episodic tremor and slip

Naum I. Gershenzon[1] and Gust Bambakidis [2]

[1]Physics Department & Department of Earth and Environmental Sciences, Wright State University, 3640 Colonel Glenn Highway, Dayton, OH 45435
[2] Physics Department, Wright State University, 3640 Colonel Glenn Highway, Dayton, OH 45435

**Abstract**

Bursts of tremor accompany a moving slip pulse in Episodic Tremor and Slip (ETS) events. The sources of this non-volcanic tremor (NVT) are largely unknown. We have developed a model describing the mechanism of NTV generation. According to this model, NTV is a reflection of resonant-type oscillations excited in a fault at certain depth ranges. From a mathematical viewpoint, tremor (phonons) and slip pulses (solitons) are two different solutions of the sine-Gordon equation describing frictional processes inside a fault. In an ETS event, a moving slip pulse generates tremor due to interaction with structural heterogeneities in a fault and to failures of small asperities. Observed tremor parameters, such as central frequency and frequency attenuation curve, are associated with fault parameters and conditions, such as elastic modulus, effective normal stress, penetration hardness and friction. Model prediction of NTV frequency content is consistent with observations. In the framework of this model it is possible to explain the complicated pattern of tremor migration, including rapid tremor propagation and reverse tremor migration. Migration along the strike direction is associated with movement of the slip pulse. Rapid tremor propagation in the slip-parallel direction is associated with movement of kinks along a 2D slip pulse. A slip pulse, pinned in some places, can fragment into several pulses, causing tremor associated with some of these pulse fragments to move opposite to the main propagation direction. The model predicts that the frequency content of tremor during an ETS event is slightly different from the frequency content of ambient tremor and tremor triggered by earthquakes.

**Introduction**

Propagation of a slip pulse along a subduction fault is accompanied by massive bursts of non-volcanic tremor (NVT). This periodic phenomenon, known as an episodic tremor and slip (ETS), has been observed in virtually all major subduction zones (Obara, 2002; 2009; Rogers, and Dragert, 2003; Kostoglodov et al., 2003; Kao et al., 2005; Rubinstein et al., 2010; Peng and Gomberg, 2010; Gonzalez-Huizar et al., 2012). NVT may also be triggered by seismic waves from large earthquakes (Obara, 2003; Rubinstein J. et al,. 2007, 2009; Peng et al., 2009;

Miyazawa and Mori, 2006; Miyazawa and Brodsky, 2008; Fry et al., 2011; Zigone et al., 2012; Chao et al, 2013) and even be modulated by tidal loading (Rubinstein J. et al., 2008; Nakata et al., 2008; Thomas et al., 2009; Lambert et al., 2009). All types of NTV, i.e. ambient tremor, tremor associated with ETS events and tremor triggered by earthquakes and tidal waves, are usually observed in the same areas, with the same frequencies and polarizations (Rubinstein et al., 2009; Peng et al., 2009; Chao et al, 2013), suggesting that they are generated by the same physical mechanism (e.g. Rubinstein et al., 2010). The suggested physical mechanisms are hydraulic fracturing (Obara, 2002; Katsumata and Kamaya, 2003; Miyazawa and Brodsky, 2008) and shear faulting (e.g. Rogers and Dragert, 2003; Shelly et al., 2007a; Miyazawa and Brodsky, 2008; Nakata et al., 2011; Ben-Zion, 2012). It is commonly assumed that NVT consists of a swarm of low frequency earthquakes (LFE) (Shelly et al. 2006; 2007a; Ide et al., 2007; Wech and Creager, 2007; La Rocca, 2009). We are developing a complementary approach, according to which LFE and NVT are manifestation/reflection of the same process, i.e. resonant-type oscillations excited inside a fault under specific conditions. Although we do not know the exact scope of these necessary conditions, the key condition for resonant-type oscillations and hence for appearance of LFE and NVT is a very low effective normal stress (Gershenzon & Bambakidis, 2014, hereafter referred to as GB). This is consistent with the fact that tremor and LFE are triggered by seismic waves from distant earthquakes, since only if the effective normal stress is small enough the small changes of shear stress associated with seismic waves can trigger a failure or a slip.

The model we have developed is based on the Frenkel-Kontorova (FK) model. It has been shown that the FK model may describe quantitatively the frictional processes between two surfaces. Accordingly it has been applied to describe laboratory frictional experiments (Gershenzon & Bambakidis, 2013), regular earthquakes (Gershenzon et al, 2009) and tremor migration patterns in ETS phenomena, as well as the scaling law of slow slip events (Gershenzon et al., 2011). The FK model itself may be described by the sine-Gordon (SG) equation, a widely used nonlinear equation of modern physics. This equation enables us to relate kinematic parameters of the frictional process, such as slip and slip velocity, with dynamic parameters such as normal and shear stress and material properties such as elastic modulus and hardness. Its basic solutions are kinks (or solitons), breathers and (anharmonic) lattice vibrations (McLauglin & Scott, 1978, hereafter referred to as MS). Kinks are stable, spatially localized, formations which can move freely in either the positive or negative direction with speeds varying from zero up to the speed of an elastic wave. Under some conditions, kinks of opposite sign can form certain stable and localized configurations known as breathers. The internal energy of a breather lies between zero and the energy of two isolated kinks. The breather energy alternates between potential energy and kinetic energy, similar to a standing wave on a spring, hence the name "breather". In our context the two types of solutions, kinks and lattice vibrations, may be interpreted as slip pulses and tremor, respectively. In the framework of the model, tremor may arise due to a variety of mechanisms, such as acceleration or deceleration of a slip pulse, interaction of a slip pulse with large asperities, and the action of an external stress disturbance on the frictional interface. We explored the latter mechanism in the first part of this article (GB). In this (second) part we focus on mechanisms describing the generation of tremor during ETS events.

Here is our scenario. During an ETS event, the accumulated shear stress in a subduction zone relaxes due to the appearance and, in contrast to a regular earthquake, the slow movement of a slip pulse. Interaction of the slip pulse with structural heterogeneities, i.e. with asperities of various sizes, emits radiation inside the fault. Additionally, the slip pulse triggers small events such as LFE by destroying some of the asperities. The triggered local failure also excites a radiation mode inside the subduction zone. This radiation (as a small-amplitude, localized relative motion of plate surfaces with zero net slip) propagates along the fault and is attenuated due to friction and geometrical spreading, resulting in S waves (tremor) propagating to the Earth's surface. So in this model tremor is a result of specific resonant-type radiation generated inside a fault. In this scenario, tremor appears and propagates in the same way as ambient tremor and tremor triggered by earthquakes (GB). However the presence of a slip pulse modifies the quantitative characteristics of tremor compared to the case of no pulse. In the following we will distinguish between tremor during ETS events ("ETS tremor") and the rest of NVT, i.e. ambient and triggered by earthquake ("ambient tremor").

The distinguishing features of tremor associated with ETS events, i.e. long duration, frequency contents slightly different compared to ambient tremor (Shelly et al, 2007a; Zhang et al, 2011; Gomberg et al, 2012), and migration pattern (Kao et al., 2006; Shelly et al., 2007b; Obara, 2009; Ghosh et al., 2010a; Ghosh et al., 2010b) (including the phenomenon of reverse migration of tremor (Houston et al, 2011)), are described by our model.

The rest of the manuscript is organized as follows. The Model describes the basics of the model. Calculations of spatial and temporal distribution of ETS tremor and its frequency content are presented in Non-Volcanic Tremor. The results are applied in the Discussion to the quantitative assessment of tremor parameters and comparison with ambient tremor. In the Conclusion we summarize the specific predictions of our model.

**The Model**

It has been shown that the dynamics of dry macroscopic frictional processes may also be described by the FK model (Gershenzon et al., 2009; Gershenzon and Bambakidis, 2013). Note that this model has been used to describe micro- and nano-scopic friction (e.g. Vanossi et al, 2004 and references therein). Other mass-spring models are widely utilized for various purposes, including the description of frictional processes (e.g. Braun et al, 2009). In addition, Burridge-Knopoff (BK)-type models are typically employed to simulate spatial and temporal patterns of seismicity (Burridge and Knopoff, 1967). The dynamics of the BK chain are defined by selecting a specific nonlinear relationship between the frictional force and the velocity. In the FK model the nonlinear behavior of the chain is implicit and does not require an explicit frictional force. The FK model has some features in common with the Toda lattice model (Toda 1989), as well as with mass-spring models describing wave propagation in a medium with periodic structure (e.g. Santosa and Symes, 1991). But from our point of view, the most adequate model for describing dry macroscopic friction in general and ETS events in particular, is the FK model.

We consider the asperities on one of the frictional surfaces (Figure 1) as forming a linear chain of balls of mass *M*, each ball interacting with its nearest neighbors on either side via spring forces of constant $K_b$ (Figures 1b and 1c). The asperities on the opposite frictional surface are regarded as being part of the rigid substrate interacting with the masses *M* via a periodic potential. Then we can apply the one-dimensional FK model to describe the slip dynamics (Kontorova and Frenkel, 1938; Hirth and Lothe, 1982):

$$M \frac{\partial^2 u_i}{\partial t^2} - K_b(u_{i+1} - 2u_i + u_{i-1}) + F_d \sin(\frac{2\pi}{b} u_i) = F(x,t) - f_{fr}(x,t,\frac{\partial u_i}{\partial t}), \tag{1}$$

where $u_i$ is the shift of ball *i* relative to its equilibrium position, *b* is a typical distance between asperities, *t* is time, $F_d$ is the amplitude of the periodic force on the mass *M* associated with the periodic substrate potential, $f_{fr}$ is the frictional (or dissipative) force, and *F* is the external (or driving) force. A list of the most important variables and parameters is presented in Table 1. Using the continuum limit approximation and expressing the coefficients *M*, $K_b$ and $F_d$ through the elastic parameters and the normal stress between frictional surfaces, we may express equation (1) in the form (Gershenzon and Bambakidis, 2013),

$$\frac{\partial^2(\frac{2\pi u}{b})}{\partial (tc/b)^2} - \frac{\partial^2(\frac{2\pi u}{b})}{\partial (x/b)^2} + A^2 \sin\left(\frac{2\pi u}{b}\right) = (F - f_{fr})\frac{2\pi A^2}{\mu b^2}, \tag{2}$$

where $c^2 = \frac{2\mu}{\rho(1-v)} = \frac{c_l^2(1-2v)}{(1-v)^2}$, $c_l$ is the longitudinal acoustic velocity (or *P* wave velocity); $\mu$ is the shear modulus and $v$ is the Poisson ratio and $\rho$ is the density. The dimensionless parameter *A* reflects how deeply the asperities from two opposing surfaces interpenetrate, and its value can be considered as the ratio between actual and nominal contact areas, hence as the ratio between the effective normal stress $\Sigma_N$ and the penetration hardness $\sigma_p$: $A \approx \Sigma_N/\sigma_p$. The equivalent form of equation (2) is the well-known sine-Gordon (SG) equation:

$$\frac{\partial^2 u}{\partial t^2} - \frac{\partial^2 u}{\partial x^2} + \sin(u) = \Sigma_S^0 - f, \tag{3}$$

where the dimensionless variables *u*, *x* and *t*, respectively, are in units of $b/(2\pi)$, *b/A* and *b/(cA)*, and the source terms $\Sigma_S^0$ and *f* are the external shear stress and frictional force per unit area, both in units of $\mu A/(2\pi)$. The variables $\varepsilon = \sigma_s = \frac{du}{dx}$ and $w = \frac{du}{dt}$ are interpreted as the dimensionless strain, stress and slip velocity in units of $A/(2\pi)$, $\mu A/\pi$ and $cA/(2\pi)$, respectively. Note that the SG equation has also been used in phenomenological models of solitary wave in the Earth's crust (Nikolaevskiy, 1996; Bykov 2001).

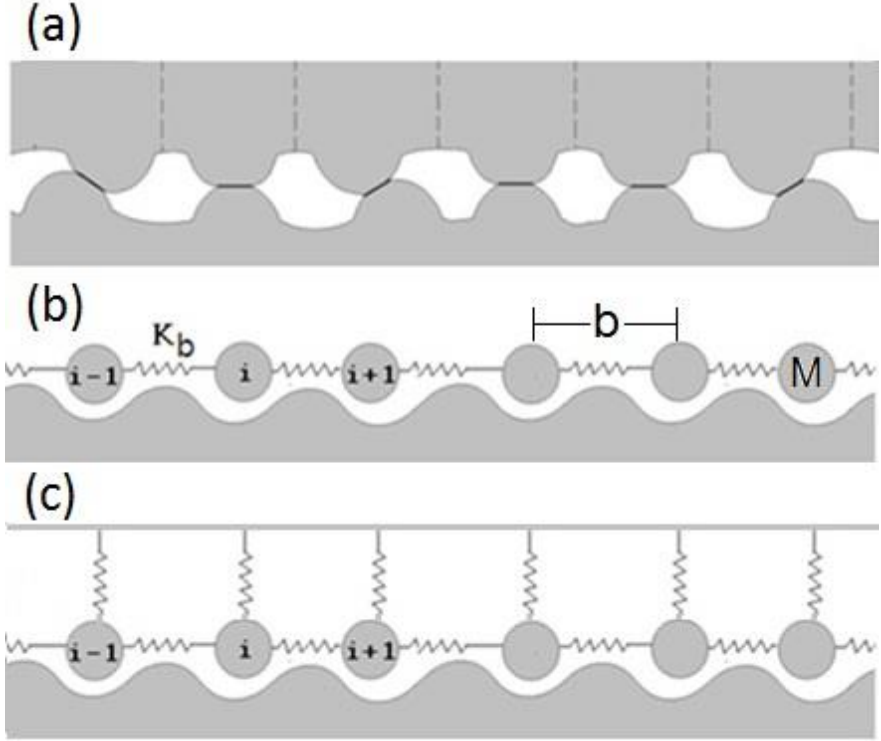

Figure 1. Schematic of asperity contact (a) and chain of masses interacting via elastic springs and placed in a periodic potential (substrate) (b) and (c). The balls represent asperities. The sine-shaped surface is the opposite plate. The horizontal and vertical harmonic springs model interaction between asperities on the same and opposite plates, respectively. In the classic FK model the harmonic forces arise due to motion of a ball along the uneven surface in a gravitational field (b), whereas in our model the harmonic forces arise due to the vertical springs (c). The mathematical descriptions of models (b) and (c) are identical.

The basic solutions of the SG equation are anharmonic vibrations (phonons), kinks (solitons), and breathers. We will use the first two solutions. Let us find the solution in the form of a simple traveling wave $u = u(x - Ut) = u(\xi)$ propagating with dimensionless velocity $U$ (scaled by $c$). If $U^2 > 1$, the solution is the phonon mode (MS):

$$u = 2\arcsin(m^{\frac{1}{2}} sn[\frac{\xi - \xi_0}{(U^2-1)^{\frac{1}{2}}}, m]),$$

where $sn$ is the Jacobi elliptic function of modulus $m$ ($0 \leq m \leq 1$) and $\xi_0$ is the initial phase. The value of $m$ is defined in terms of the wave amplitude $a$ by the relation: $m = [\frac{1}{2}(1 - \cos a)]^{0.5}$. If $U^2 < 1$, the solution is the solitonic (localized wave) solution, described by the two-parameter formula (MS):

$$u_0(x, t; U, X) = 4tan^{-1}\exp[\pm\left(\frac{x - \int_0^t U dt' - X}{(1-U^2)^{0.5}}\right)], \qquad (4)$$

where $X$ is the initial position of the center of the soliton. Perturbations such as an external force, friction and structural heterogeneities affect the movement of solitons. Thus interaction between these two modes is possible if the terms on the right-hand side of equation (3) are not zero. The respective mathematical apparatus for small perturbations has been developed by McLauglin and Scott (MC) and we will use their approach here (see Appendix).

The nonlinear dispersion equation for the phonon mode is (MS):

$$\omega^2 - k^2 = \frac{\pi^2}{4K^2(m)}, \qquad (5)$$

where ω is the angular frequency in units of $cA/b$, $k$ is the wave number in units of $A/b$, and $K$ is the complete elliptic integral of the first kind. If the amplitude is large, i.e. $a \lesssim b/2$ (hence if $m \lesssim 1$) the right hand side of the dispersion relation (5) is nullified and (5) reduces to $\omega = k$ (see the bottom curve in Figure 3 from (GB)). If the wave amplitude is small, i.e. $a \ll \frac{b}{2}$ ($m \ll 1$), the dispersion relation (5) is simplified to (see the top curve in Figure 3 of (GB)):

$$\omega^2 - k^2 = 1. \qquad (6)$$

So for waves with amplitude much less than $b/2$ (dimensionless amplitude $a \ll \pi$), the dimensionless group velocity $V$ of the wave packet (in units of $c$) is:

$$V = \frac{d\omega}{dk} = \frac{k}{(1+k^2)^{0.5}}. \qquad (7)$$

As already pointed out (see GB), two important conclusions follow from equations (6) and (7): (1) the angular frequency cannot be less than unity (ω≥1), which means that the minimal frequency of the disturbance generated inside a fault, $f_{min} = \left(\frac{cA}{b}\right)\left(\frac{1}{2\pi}\right) = \left(\frac{cA}{2\pi b}\right)$ in dimensional units, is defined by the fault parameters and effective normal stress only and does not depend on the parameters of a particular source; (2) the group velocity of a disturbance propagation along a fault may vary from values much less than $c$ up to the value $c$. The frequency $f_{min}$ has significance as a quasi-resonant frequency.

**Non-Volcanic Tremor**

Collisions of the slip pulse with small structural heterogeneities on the frictional surface causes the emission of radiation (tremor). Note that this radiation is not associated with any mechanical failure. However the movement of a slip pulse may trigger small failures, i.e. LFE and/or VLF, in the same way as seismic waves from distant earthquakes do. The LFEs are part of tremor and VLF events trigger tremor. To model these processes we will assume the presence of small localized structural heterogeneities (hence asperities of different sizes) or a small localized failures which can be long in time or short in time. Thus we can distinguish between three sources of tremor: 1) fast failure such as a regular earthquake or LFE in the presence of a slip pulse, 2) slow failure such as a VLF earthquake or aseismic slip and 3) interaction of a slip pulse with a structural heterogeneity.

*Fast initial failure*

Let us model the spontaneous failure within a fault as an impulse localized in space and time. Using this as the source term in equation (3), i.e. $\Sigma_S^0 - f = a_{init}\delta(x)\delta(t)$, where $\delta$ is the Dirac delta-function and $a_{init}$ is the strength of the disturbance, we look for the phonon radiation field inside the fault produced by this impulse in the presence of a slip pulse. Results of a perturbation analysis of the SG equation (MS) gives the "slip field" produced by the perturbation (see formula (A4) from Appendix). Figure 2 shows the results of a numerical integration of the temporal and spatial distribution of slip $u$ for $a_{init} = 1$. One can see that the disturbance propagates along a fault in both directions with unit velocity (velocity *c* in dimensional units) (Figure 2 left panel). The wave number *k* ranges from large values close to the wave fronts to small values close to the center. The value of *k* at the center decreases in time and becomes much less than unity when *t*>>2π. In this case *ω*≈1 (see equation (6) and Figure 2b), so after a short time the frequency of the oscillation in close proximity to the center reaches the value *f*=*ω*/(2π)≈1/(2π) and does not change much thereafter. The oscillation frequency at points close to the fronts is higher (see Figure 2c and 2e). Therefore the angular frequency ranges from a low of *ω*=1 at the center to higher frequencies $\omega = (1 + k_{init}^2)^{0.5}$ at the periphery of the disturbance, where $k_{init}$ is the characteristic wave number of the initial disturbance (see Figures 2 right panel). This behavior is similar to the analogous fast initial failure case considered in part I of this article (GB). However these two cases are not identical and the differences is visible (compare Figures 2 from this article and Figure 4 from (GB)), suggesting that the frequency content of tremor generated by the same mechanism under the same conditions but in the presence of a slip pulse is different from when the slip pulse is absent. Figure 3 shows the associated spatial and temporal distributions of slip velocity $w(x,t) \equiv \frac{du}{dt}$ and shear stress $\sigma_s(x,t) \equiv \frac{du}{dx}$.

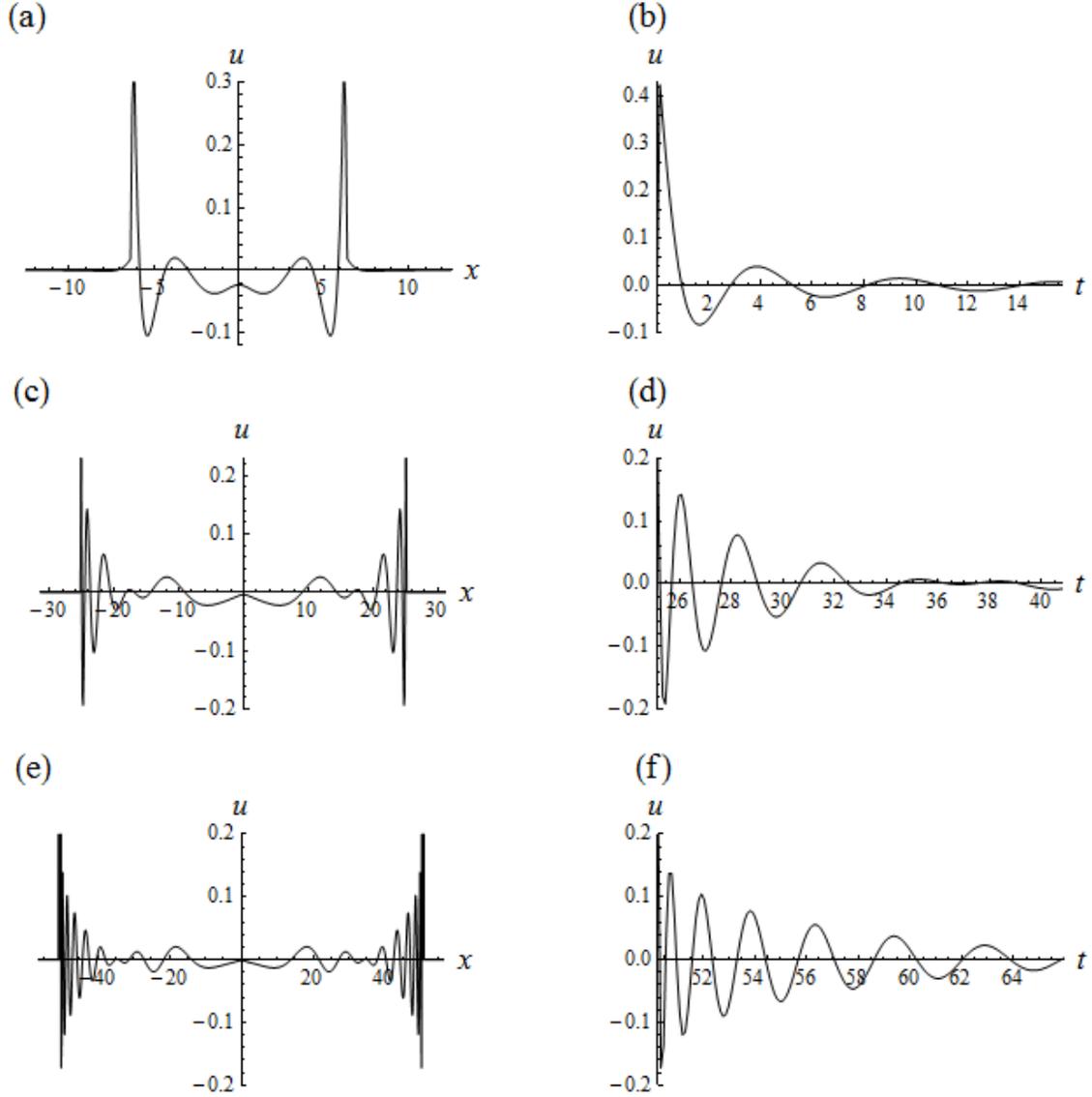

Figure 2. Evolution of the shift disturbance in space (along a fault) for various times ($t = 2\pi$ (a), $t = 8\pi$ (c), and $t = 16\pi$ (e)) and in time for various distances from the center ($x = 0$ (b), $x = 8\pi$ (d), and $x = 16\pi$ (f)). Disturbance originates at point $x=0$ and time $t=0$ due to the external source $\delta(x)\cdot\delta(t)$ and propagates in both directions. Center of slip pulse is at $x=0$. Local wave number at any particular time decreases going from the center to the wave fronts (Figure 2(e)). Period of oscillation at point $x=0$ approaches $2\pi$ ($f\approx 1/(2\pi)$) after a short time ($t>2\pi$) from the beginning (see Figure 2(b)). Oscillation period at points $x=8\pi$ and $16\pi$ progressively decreases compared with the value at point $x=0$ (see Figure 2(c-f)). Disturbance at the center plays the dominant role in the frequency distribution of the emitted tremor; peripheral disturbance contributes to its high frequency range. Here and in all subsequent figures the variables on both axes are dimensionless.

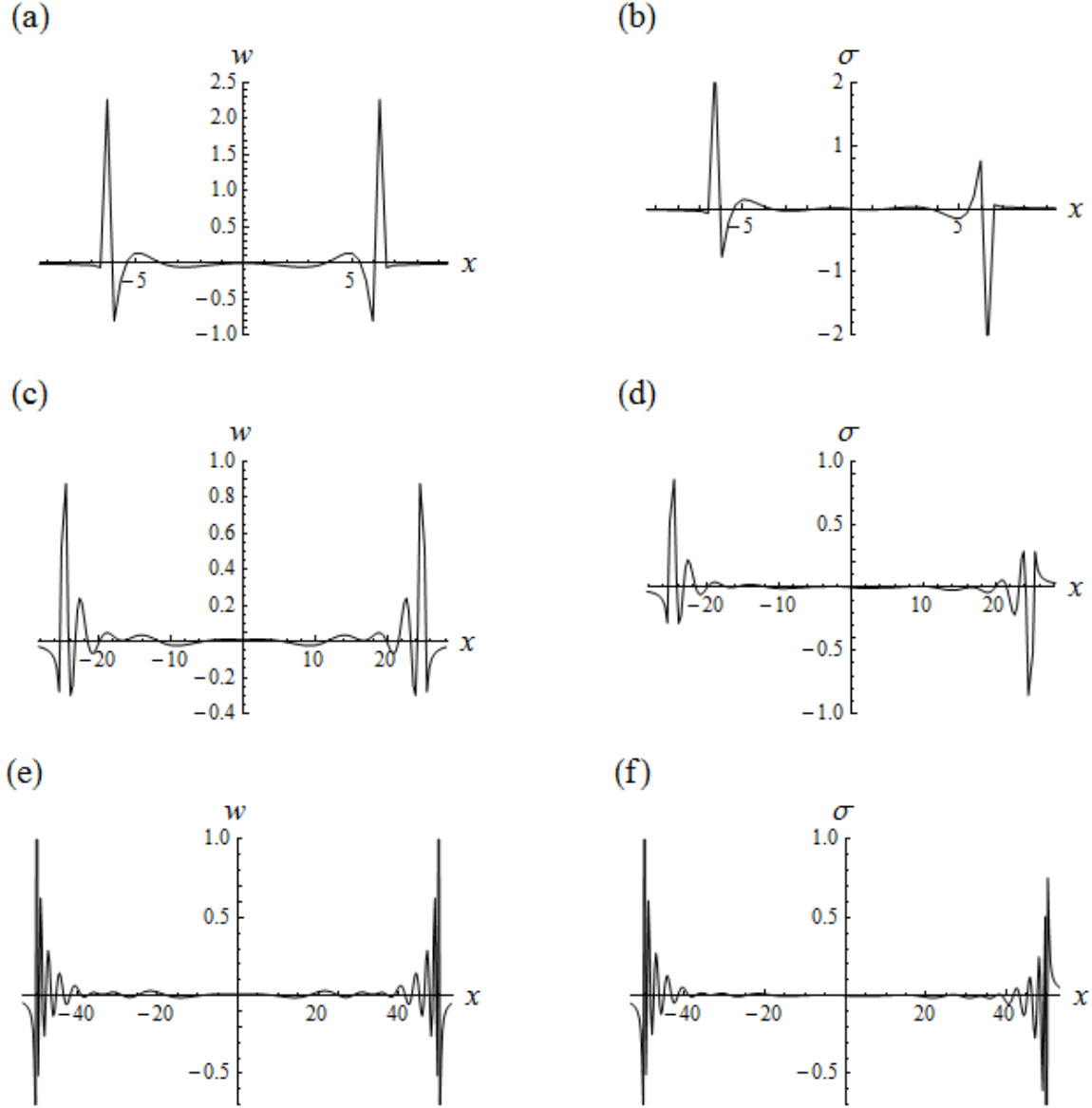

Figure 3. Spatial distribution of slip velocity $w$ (panels on the left) and shear stress $\sigma$ (panels on the right) along a fault for various times (($t = 2\pi$ for (a) and (b), $t = 8\pi$ for (c) and (d), and $t = 16\pi$ for (e) and (f)). Disturbance originates at point $x=0$ and time $t=0$ by the external source $\delta(x)\cdot\delta(t)$ and propagates in both directions.

To consider a realistic case we introduce friction together with initial and boundary conditions. Let us solve equation (3) with the right hand side of the form:

$$\Sigma_S^0 - f = a_{init} \left(\frac{9}{\pi}\right)^{0.5} exp(-9x^2)\delta(t) - [\alpha_s sign(u_t) + \alpha_d u_t], \tag{8}$$

where the first term on the right hand side represents the initial conditions that the integral over $x$ equal $a_{init}$, the second term is the external shear stress, and the last term denotes friction. As in GB, we introduce friction as a sum of two terms: the first term represents static friction and the second represents viscous damping, in their generally accepted form. The boundary conditions are $u(x=x_-)=u(x=x_+)=0$ and $u_t(x=x_-)=u_t(x=x_+)=0$, where $x_-$ and $x_+$ are the left and right positions of the boundary. Figure 4 shows the evolution of a signal in time and space in the presence of the slip pulse described by equation (4) with $U=0$ and $X=0$. As one can see at Figure 4, the resulting disturbance consists of two parts: small amplitude phonons propagating with unit velocity in both directions and the shift of the slip pulse from its initial position. So the impulse-like failure not only generates phonons but also shifts the slip pulse.

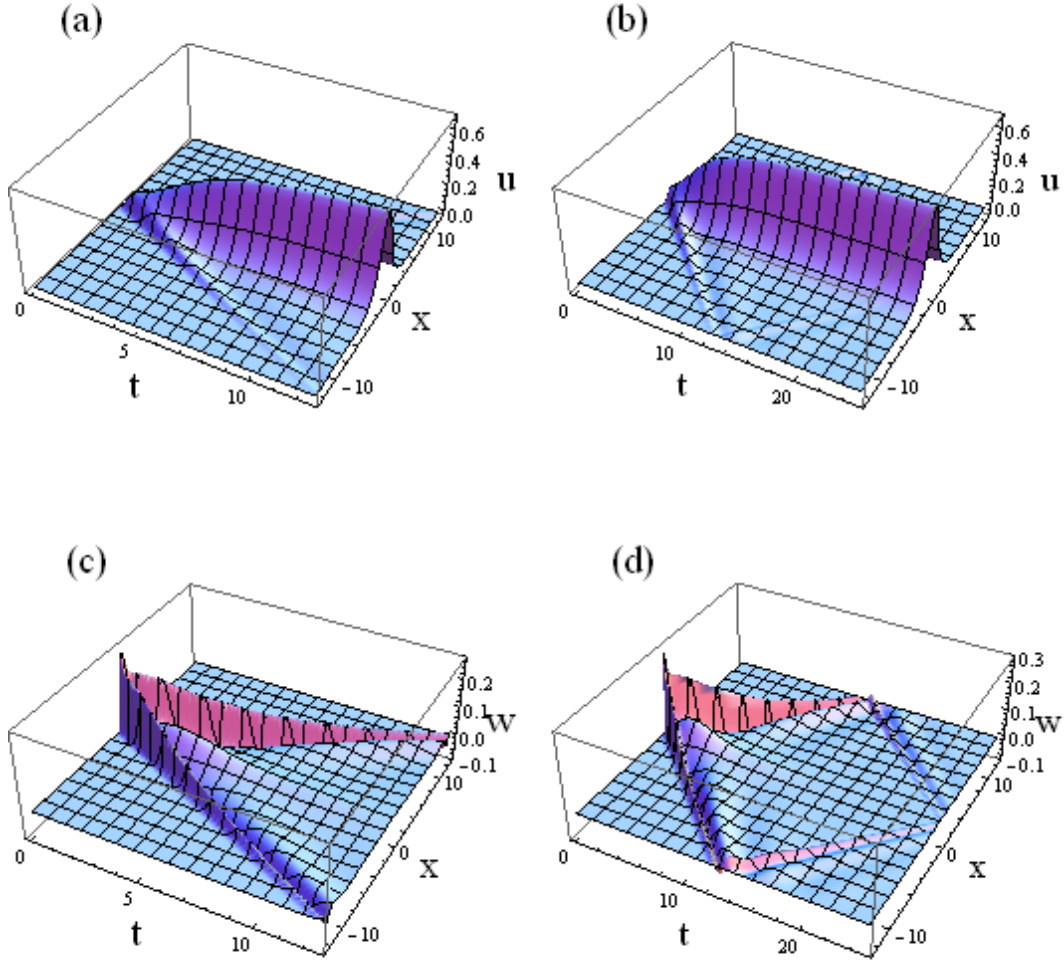

Figure 4. Temporal and spatial evolution of a signal (slip $u$ and slip velocity $w$) inside a fault for different values of boundary position ($u(x = -16\pi) = u(x = 16\pi) = 0$ for (a) and (c) and $u(x = -4\pi) = u(x = 4\pi) = 0$ for (b) and (d), computed for the fast initial failure, i.e. for LFE. Computation made with $\alpha_s = \alpha_d = 0.025$. Center of a slip pulse at $x=0$. Disturbance consists of two parts: (1) small amplitude phonons propagating with unity velocity in both directions; (2) the shift of a slip pulse. Note that the pulse width is about $2\pi$, which explains the width of the central object around position $x=0$.

Since the plate surface is part of the earth's crust, its periodic localized oscillations will generate *S*-type seismic waves with the same mix of frequencies; the latter may propagate through the crust to the Earth's surface. To determine the disturbances (tremor) at a given point $(x_s, y_s, z_s)$ on the Earth's surface produced by the disturbance generated inside a fault, we integrate velocity *w* over the entire source. In the far zone approximation, the expression for calculating the Earth's surface shift, $u_s$, is (Aki & Richards, 1980):

$$u_s(x_s, y_s, z_s, t) = \int_0^t \int_{x_-}^{x_+} \int_{y_-}^{y_+} \frac{gw(x, t' - r/c_s)}{r} dx dy \, dt', \qquad (9)$$

where $y_-$ and $y_+$ form the boundary of the source in the *y* direction, *r* is the distance between source and observation point, $c_s$ is the shear wave velocity, and *g* is a coefficient. To find $u_s$, we first integrate equation (3) to obtain *u* and hence $w = du/dt$. Inserting the latter into equation (9) we find $u_s$ after integration over the *x* and *y* coordinates and time *t'*. Figure 5 shows the results of a numerical integration of the velocity $v = du_s/dt$, as well as the spectral content for various values of friction coefficients. The position of the center of a slip pulse coincides with the position of an impulse-like failure. We can see that increasing the friction coefficients decreases (obviously) the signal life-time and increases the spectral amplitude at higher frequencies. The dashed lines show the spectral density of the same signal but including a frequency-dependent attenuation during propagation from the source to the measurement point. To calculate the attenuation we multiply the spectrum by $\exp(-\pi ft/Q)$, where the travel time $t = 15$s and $Q = 350$ (using the same approach as Beroza and Ide (2011)). Note that in Figure 5 we display the frequency content in the interval from $1/(4\pi)$ to $30/(2\pi)$ (hence from 0.5 to 30 Hz in dimensional units) to show the maximum at frequency $1/(2\pi)$ (or 1 Hz).

Interaction between phonons and pulse modifies the frequency content of a signal initiated by an impulse-like failure (compare Figure 5 from this article and Figure 8 from (GB)). Figure 6 illustrates the dependence of the spectrum on the distance between pulse and failure point.

*Slow initial failure*

Now consider the case where the source term in equation (3) has the form $\Sigma_S^0 - f = a_{init} \delta(x) \eta(t)$, where η is the Heaviside function. As before, using perturbation analysis we find the expression for $u(x, t)$ (see Appendix formula (A5)) and associated slip velocity (friction is not included). Figure 7 shows the results of a numerical integration. From this figure one can see that, although the shape of the oscillations is different from the case of fast initial failure (compare Figures 2 with 7), the main features are the same, namely 1) fronts propagate with unit velocity, 2) the disturbance includes small *k* at the center and large *k* close to the boundaries, and 3) the frequency of oscillation in the central area is close to $2\pi$.

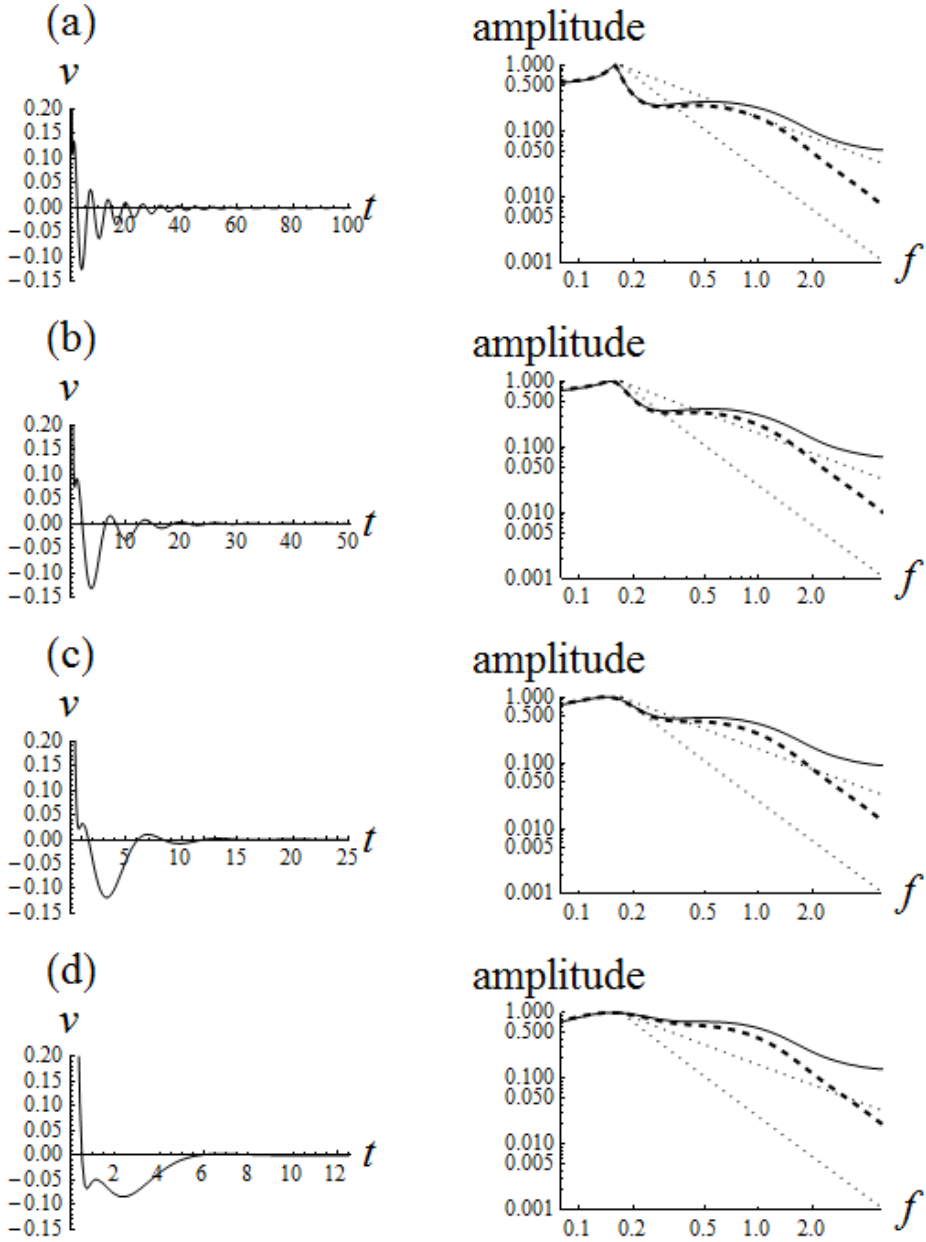

Figure 5. Calculated tremor, i.e. velocity of Earth's surface movement (in arbitrary units), versus time (in dimensionless units) produced by the "fast" disturbances for various values of friction coefficients ($\alpha_s = \alpha_d = 0.01, 0.025, 0.05$, and $0.1$ for (a), (b), (c), and (d), respectively). Also shown is corresponding frequency content of tremor in range from $1/(4\pi)$ to $30/(2\pi)$ (hence from 0.5 to 30 Hz in dimensional units). Reference fall-offs (dotted line) are $f^{-1}$ and $f^{-2}$. Dashed lines show the spectral density of the same signal but taking into account frequency-dependent attenuation during its propagation from the source to the measurement point. To calculate attenuation we multiply the spectrum by $\exp(-\pi f t/Q)$, where the travel time $t = 15$ s and $Q = 350$ (Beroza and Ide, 2011).

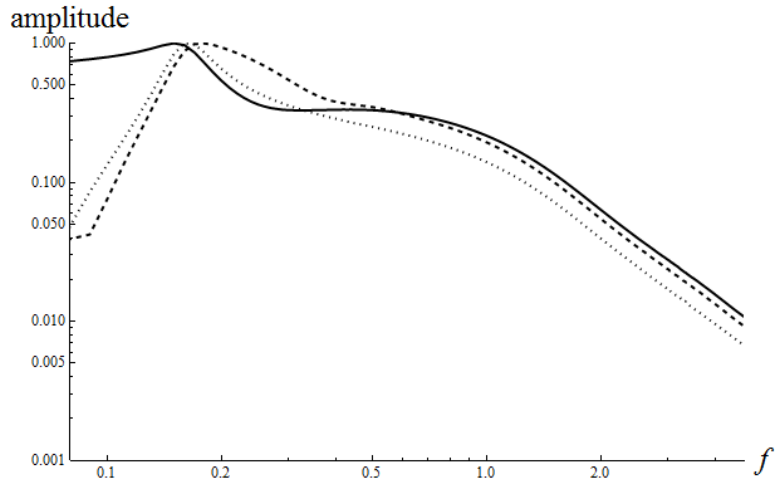

Figure 6. Frequency content of tremor adjusted by frequency-dependent attenuation. Calculation made for various distances $d$ between position of an impulse-like failure and position of the center of slip pulse: $d = 0$ (solid line), $d = \pi$ (dashed line) and $d = \infty$ (dotted line). Friction coefficients $\alpha_s = \alpha_d = 0.025$.

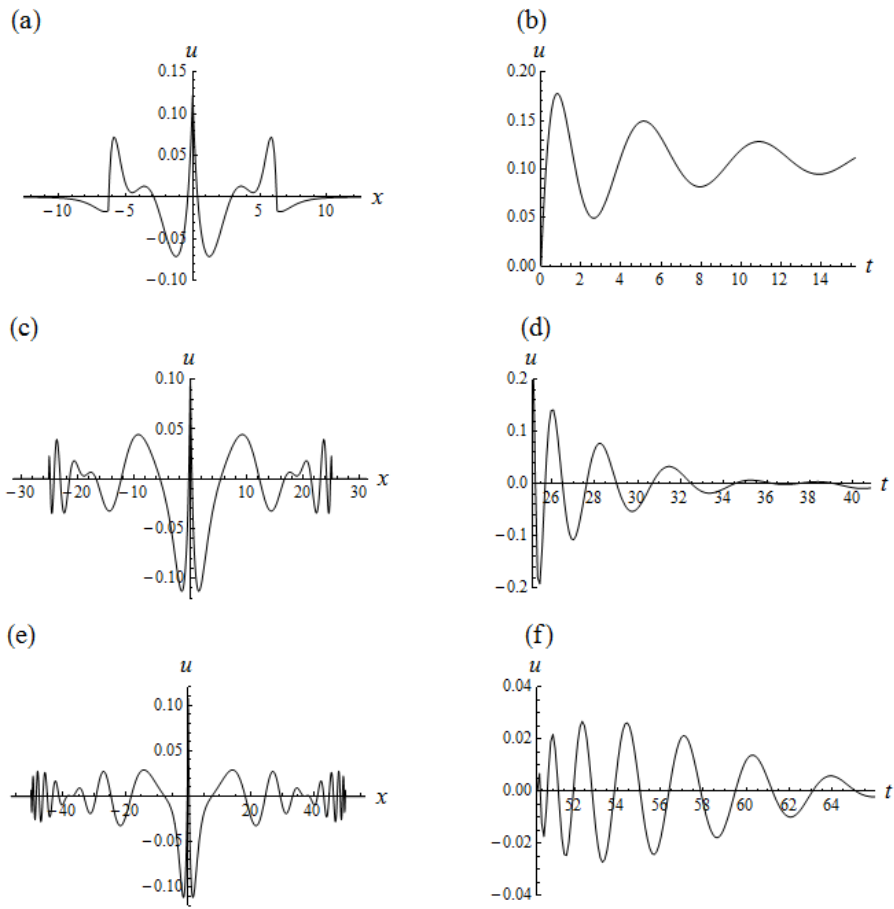

Figure 7. The same as in Figure 2 for the external source $\delta(x)\cdot\eta(t)$.

To include friction we solve equation (3) numerically with the right hand side in the form:

$$\Sigma_S^0 - f = a_{init} \left(\frac{3}{\pi}\right)^{0.5} exp(-3x^2)\eta(t) - [\alpha_s sign(u_t) + \alpha_d u_t], \qquad (10)$$

The spatial and temporal evolution of the signal is shown in Figure 8 and its spectral composition in Figure 9. Note that, although the central frequency of tremor initiated by the fast and slow failures is the same, the spectral compositions are noticeably different, i.e. the spectral fall-off is larger for the tremor initiated by slow initial failure for the same values of friction coefficients (compare Figure 5 and 9).

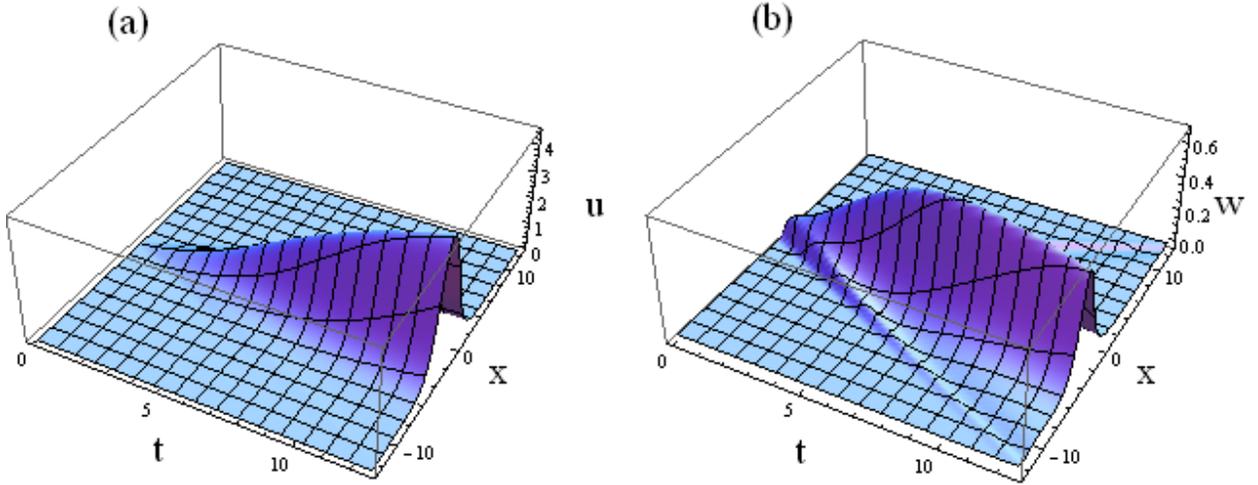

Figure 8. Temporal and spatial evolution of a signal, (a) slip $u$ and (b) slip velocity $w$, inside for a VLF-like event. Computation made with $\alpha_s = \alpha_d = 0.025$. The center of slip pulse is at $x=0$.

### *Interaction between slip pulse and structural heterogeneity*

Now we consider tremor emission due to interaction of a moving pulse with structural heterogeneities. Let us rewrite equation (3) in the form:

$$\frac{\partial^2 u}{\partial t^2} - \frac{\partial^2 u}{\partial x^2} + (1 + \beta(x))\sin((1 + \varphi(x))u) = \sigma_0 - [\alpha_s sign(u_t) + \alpha_d u_t], \qquad (11)$$

where $\sigma_0$ is the external stress and $\beta(x)$ and $\varphi(x)$ are spatially localized functions modelling structural heterogeneities, i.e. asperities with larger or smaller sizes both normal ($|\beta| > 0$) and parallel ($|\varphi| > 0$) to the frictional surface (recall that equation (3) describes the case where all asperities have the same size, i.e. an "ideal" substrate). External stress pushes a slip pulse along a fault. In the case of an ideal substrate ($\beta = 0, \varphi = 0$), the pulse propagates with constant velocity. The imbalance between friction and external stress defines the velocity magnitude. Figure 10a illustrates the moving slip pulse for an ideal substrate. Figure 10b shows the dynamics of a pulse calculated for the following parameters: $\Sigma_S^0 = 0.01$, $\alpha_s = 0.01$, $\alpha_d = 0.01$, $\beta = 0$ and $\varphi = 0.2exp(-(x + 2\pi)^8)$. One can see that when the pulse reaches the heterogeneity it begins to oscillate around this spot, intensively emitting tremor.

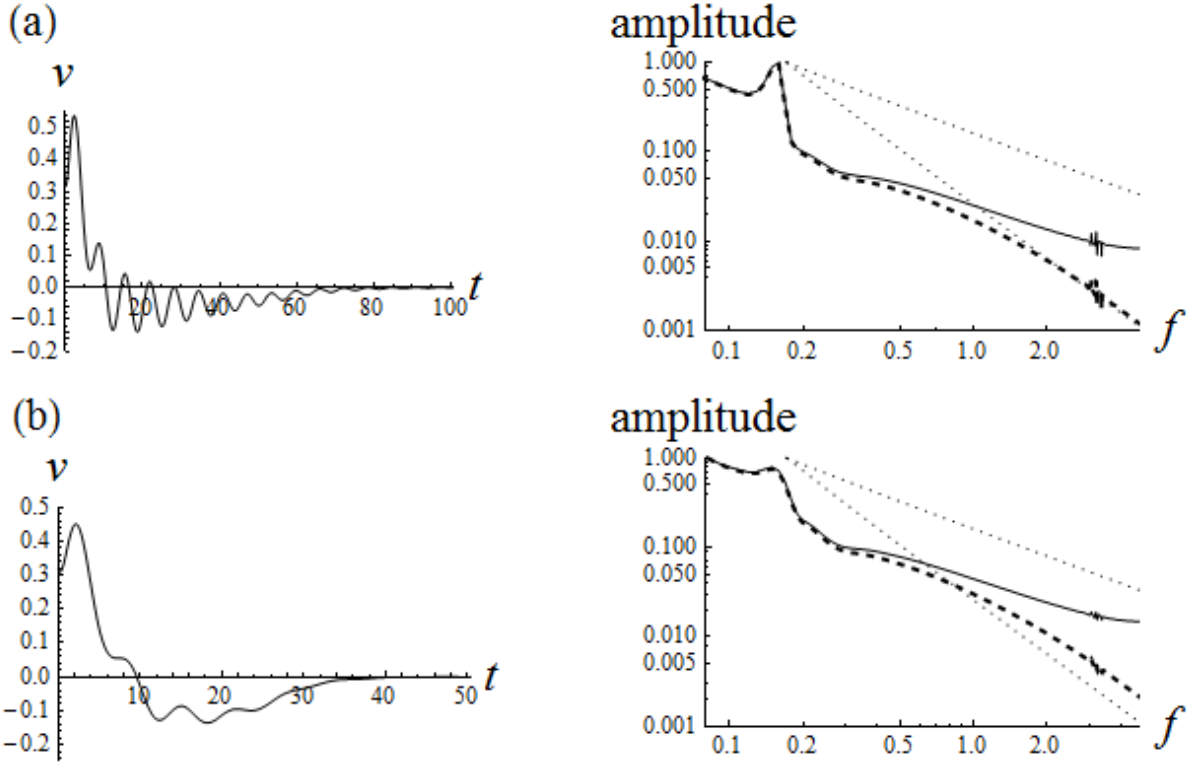

Figure 9. Calculated tremor versus time produced by "slow" disturbances for various values of friction coefficients ($\alpha_s = \alpha_d = 0.01 \; and \; 0.025$ for (a) and (b), respectively). Also shown is the corresponding frequency content of tremor in the range from $1/(4\pi)$ to $30/(2\pi)$ (hence from 0.5 to 30 Hz in dimensional units). The reference fall-offs (dotted line) are $f^{-1}$ and $f^{-2}$. Dashed lines show spectral density of the same signal but taking into account frequency-dependent attenuation during propagation from the source to the measurement point.

Inserting the solution of equation (11) into equation (9) we find the displacement and velocity of the ground at an observation point. Figure 11 depicts the results of a calculation of velocity and spectral density for two different frictional coefficients with $\beta(x) = 0.1exp(-(x - \pi/4)^2)$ and $\varphi = 0$.

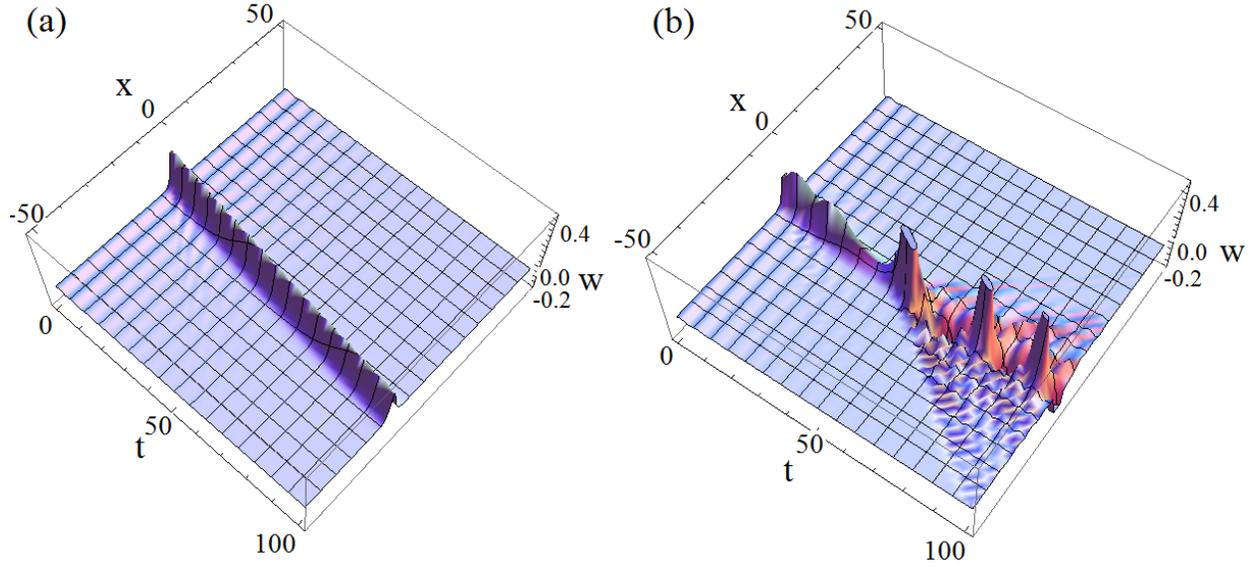

Figure 10. The slip velocity *w* of a slip pulse in time-space (*x-t*) coordinates moving in (a) ideal substrate and (b) substrate with a structural heterogeneity. Pulse is driven by constant external shear stress. Figure 10(b) shows that the pulse oscillates about an obstacle, emitting tremor. Note that a wave which we interpreted as tremor could simply reflect the instability of the calculation. To verify that this was not the case, we changed the grid size by two orders of magnitude and obtained similar results.

**Discussion**

Our model contains two adjustable parameters: the typical distance *b* between asperities and the dimensionless parameter *A*. If we suppose that a slip pulse in an ETS event is represented as a solitonic solution of equation (3) then the parameter *b* should be the typical slip produced by one ETS event, i.e. *b*≈30 mm (Rogers and Dragert, 2003). The parameter *A* is the ratio between the effective normal stress and the penetration hardness (Gershenzon and Bambakidis, 2013). We can estimate *A* assuming that the predicted minimal frequency of tremor, $f = cA/(2\pi b)$ (in dimensional units), corresponds to the lowest frequency of the observed velocity spectrum of tremor, which is about 1 Hz. Thus we find that $A \approx 4 \cdot 10^{-5}$ if *f*=1 Hz, *b*=30 mm and *c*=5 km/s (GB). Now we can estimate the typical width of the pulse, $d = \frac{2\pi b}{A} \approx 5$ km, and the number of asperities, $N = \frac{2\pi b}{A} \approx 1.6 \cdot 10^5$, occupied by a pulse. It is also of interest to estimate the pulse velocity *U* during an ETS event. Supposing that $U_{real} = 10\ km/day$ we find the dimensionless value $U = \frac{U_{real}}{c} \approx 2 \cdot 10^{-5}$.

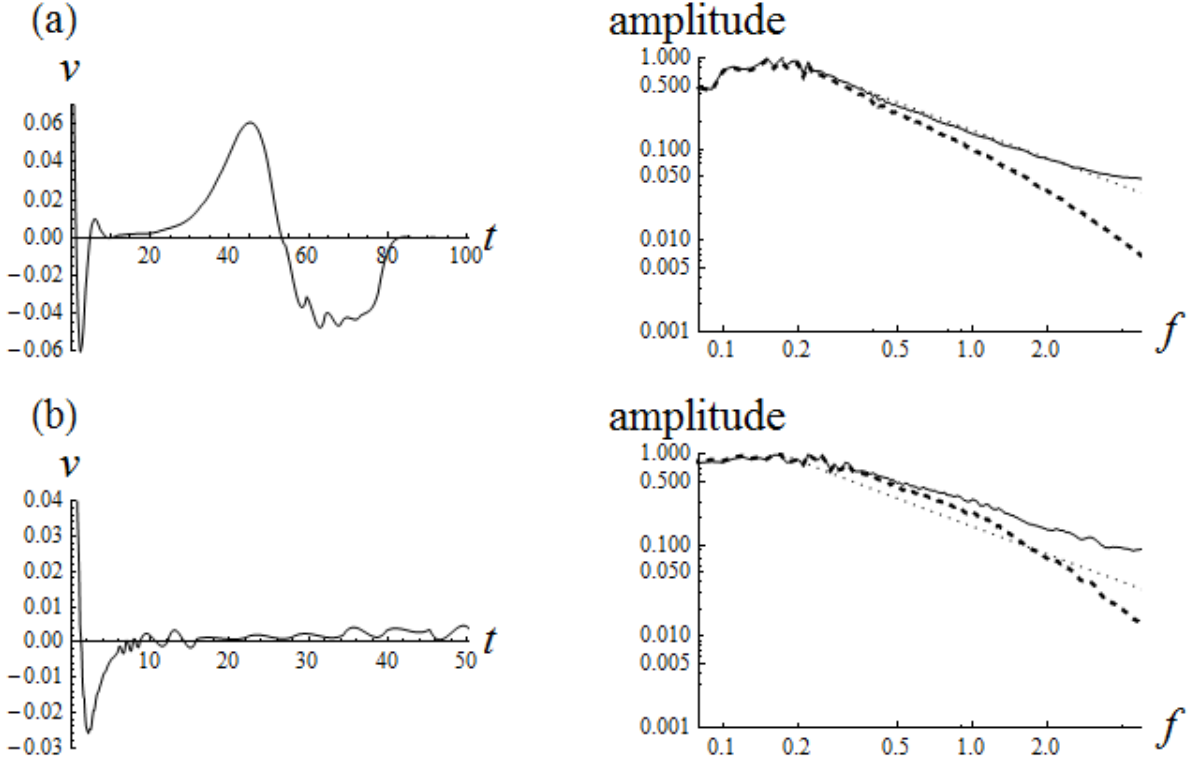

Figure 11. Calculated tremor versus time, produced by interaction of a moving slip pulse with a substrate heterogeneity, for two different values of friction coefficients ($\alpha_s = \alpha_d = 0.01$ and $0.025$ for (a) and (b), respectively). Also shown is the corresponding frequency content of tremor from $1/(4\pi)$ to $30/(2\pi)$. Reference fall-offs (dotted line) are $f^{-1}$. Dashed lines show the spectral density of the same signal but taking into account frequency-dependent attenuation during its propagation from the source to the measurement point.

The amplitude of strain $\varepsilon_a$ and shear stress $\sigma_a$ associated with a pulse are $\varepsilon_a = \frac{A}{2\pi} \approx 6 \cdot 10^{-6}$ and $\sigma_a = \frac{\mu A}{2\pi} \approx 0.2$ MPa (for µ=30 GPa). These values of strain and shear stress are large enough to trigger small failures. The latter initiates resonant-type oscillations inside a fault which are seen as tremor on the surface of the Earth. Interaction of a pulse with heterogeneities (asperities of different sizes) also may generate tremor without asperity failure, which may be the dominant mechanism of tremor generation during ETS events. This effect is analogous to the phenomenon of acoustic emission during plastic deformation of crystals (Kaiser, 1953). Whatever the source of tremor during ETS events (failure triggered by a pulse or interaction of a slip pulse with a heterogeneous substrate), the duration of a tremor swarm and its migration pattern are associated with pulse dynamics. Indeed the variety of migration patterns, such as slip-parallel long-term tremor migration along the strike direction with a velocity of 10 km/day (Kao et al., 2006; Ghosh et al., 2010a) and short-term tremor migration with a velocity of 50 km/day in both dip-up and dip-down directions (Ghosh et al., 2010b), might be explained if we considered the trajectory and velocity of a pulse itself and kinks propagating along a pulse (see Figure 2 in the article by Gershenzon et al, (2011)). The latter is analogous to the propagation of kinks moving along a dislocation in a crystal.

It has been shown (Houston et al, 2011) that tremor can "migrate rapidly back, away from the region where tremor and slip are advancing, through parts of the plate interface that have just ruptured". Figure 12 shows schematically how this phenomenon can be described in the framework of our model. Continuing the analogy between a dislocation in a crystal and a slip pulse in an ETS event, suppose that there is an obstacle (pinning point) preventing slip at some location. Parts of the pulse from the left and right sides of the obstacle continue to move, increasing tension around the obstacle (Figures 12 (b, c and d)). Eventually the slip pulse splits into two parts: one part continues to move in the original direction and the second part collapses to the obstacle (Figure 12 (e and f)). Thus, starting from the moment of pulse defragmentation, tremor would migrate in both directions, in accordance with observations (Houston et al, 2011).

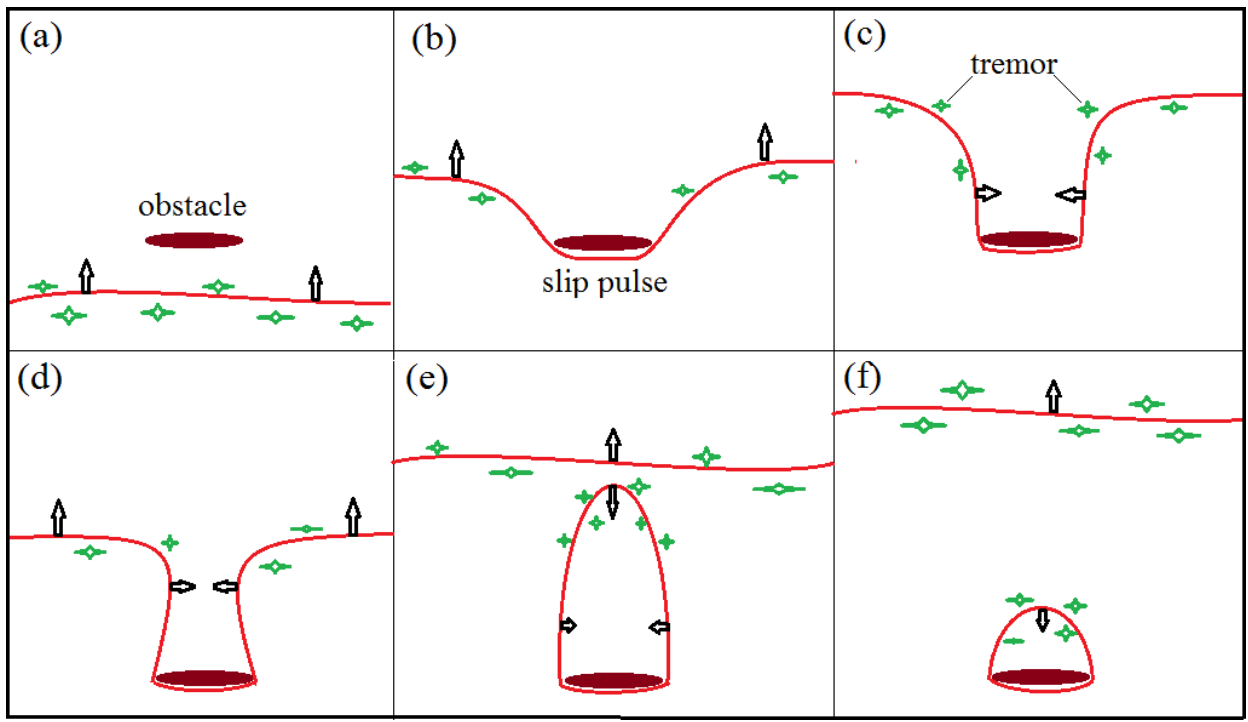

Figure 12. Schematics describing the phenomenon of reverse migration of tremor (Houston et al, 2011). Consecutive positions of a slip pulse are shown by the thick line. Arrows show the direction of pulse propagation. Stars designate tremor triggered by the moving pulse. The obstacle prevents pulse propagation through it, causing deformation of the pulse and eventually its fragmentation. The part of the pulse associated with the obstacle collapse. As a result, this part moves backward, so the tremor associated with it migrates opposite to the original direction.

In a previous article (GB) we estimated the value of the effective normal stress: $\sigma_N \approx A\sigma_p \approx 25$ kPa, where $\sigma_p \approx 0.018\mu(1+)\nu$ (Rabinowicz, 1965) and $\nu$=0.3. This estimate confirms the requirement of low effective normal stress (hence high fluid pressure) for the model to work. In particular it justifies the possibility of oscillations and/or /radiation inside a fault (see GB for a more detailed discussion), which is key to the model considered.

There is a small difference between tremor generated by a moving slip pulse (one-soliton approximation, according to MS) and ambient tremor (zero-soliton approximation). Compare Figure 5 of this article and Figure 8 of GB. In the former case the spectrum is flatter from both sides of the central frequency ($1/(2\pi)$ in dimensionless units). This difference could be used to distinguish between these two types of tremor. Tremor from ETS events should include less LFE than ambient tremor and tremor triggered by large earthquakes.

**Conclusions**

In our model, a moving slip pulse generates tremor due to small-scale failures (LFEs and VLF events) as well as by interaction of a pulse with structural heterogeneities. The latter mechanism is likely dominant in ETS events. In particular it is consistent with the observation that tremor location coincides with positions where the rate of fault slip is high (Bartlow et al., 2011). The tremor spectrum we obtain, i.e. in Figures 5(b, c and d) and Figure 11(b), is consistent with the spectrum observed in ETS events, i.e. Figure 5(b) from Rubinstein et al (2010) and Figure 3(b) from Beroza and Ide (2011). By considering the frequency-dependent attenuation of a signal coming from the typical depth of a NVT source, the latter authors concluded that the spectrum of the "original" signal, i.e. signal without attenuation, should be almost flat. In particular it means that the *v-t* curve of such a signal should have a nearly δ-function shape with a few seconds typical lifetime. This could only be the case when oscillations developed inside a fault quickly relax, such as in Figures 5(b, c and d) and 15(b). Thus, we may conclude that the observed tremor, with duration ranging from minutes to hours to days, should include multiple short-time events coming from different places. This continuous generation of tremor is expected if a slip pulse is moving along a fault. This may be the main reason why it is difficult to identify precise tremor location during an ETS event.

In the framework of the model considered, it is possible to explain the complicated pattern of tremor migration during an ETS event (e. g. Shelly et al., 2007a, Obara, 2009; Ghosh et al., 2010(a); 2010(b); Houston et al, 2011). In the work just cited, by Ghosh et al., two migration velocities, differing in both magnitude and direction, were observed by the dense seismic array installed directly above an ETS zone. Using the analogy between dislocations in crystals and slip pulses in a subduction zone, we describe the observed tremor migration as follows. The movement along the strike direction with a velocity of 10 km/day is associated with movement of 2D slip pulse in a direction normal to the pulse. The rapid tremor propagation in the slip-parallel direction with a velocity of ~50 km/hr is associated with movement of kinks along a slip pulse (see Figure 2 from Gershenzon et al. (2011) for more details). Finally, the phenomenon of reverse tremor migration may also be described using the analogy with dislocations in crystals. A slip pulse, pinned in some places, is divided into several pulse segments as depicted in Figure 12. This causes tremor associated with a pinned segment of the pulse to move opposite to the main propagation direction.

Table 1: List of main variables and parameters

| Variable | Definition | Variable | Definition |
|---|---|---|---|
| $\mu$ | shear modulus | $b$ | distance between asperities |
| $\nu$ | Poisson ratio | $\omega$ | angular frequency |
| $\rho$ | density | $f$ | frequency |
| $\sigma_p$ | penetration hardness | $k$ | wave number |
| $\Sigma_N$ | effective normal stress | $u$ | slip |
| $K_b$ | spring constant | $u_s$ | shift at the earth's surface |
| $F_d$ | amplitude of the periodic force | $w$ | slip velocity |
| $F$ | external force | $v$ | velocity at the earth's surface |
| $M$ | mass of ball/asperity | $U$ | phase velocity |
| $f_{fr}$ | frictional force | $V$ | wave packet velocity |
| $\varepsilon$ | strain | $\omega$ | angular frequency |
| $\sigma_s$ | shear stress | $f$ | frequency |
| $c_s$ | shear wave velocity | $k$ | wave number |
| $c_l$ | longitudinal vave velocity | $d$ | pulse width |
| $c$ | effective vave velocity | $a$ | wave amplitude |
| | | $A$ | $\Sigma_N/\sigma_p$ |

**Acknowledgement**

This work was supported by NSF grant EAR-1113578. Mathematica 9.0.1.1 has been used for the computations shown in Figures 2-11.**References**

Aki, K. and P.G. Richards (1980), Quantitative Seismology: Theory and Methods (part 1). Freeman and company, San Francisco, 557.
Bartlow, N. M., S. Miyazaki, A. M. Bradley, and P. Segall (2011), Space-time correlation of slip and tremor during the 2009 Cascadia slow slip event, Geophys. Res. Lett., 38, L18309, doi:10.1029/2011GL048714.
Ben-Zion, Y. (2012), Episodic tremor and slip on a frictional interface with critical zero weakening in elastic solid, *Geophys. J. Int.* 189, 1159–1168, doi: 10.1111/j.1365-246X.2012.05422.x.20

**Appendix**

The solutions of the perturbed SG equation (3) can be obtained in analytical form only in the case of a small perturbation $f$ ($f \ll 1$). We will use the results of perturbation analysis developed by *McLauglin & Scott* [1978] (hereafter MS). We suppose that there is a soliton (kink) moving along *x* direction with velocity *U*. Such a localized wave (undisturbed soliton) can be described by a two-parameter formula (MS):

$$u_0(x,t;U,X) = 4\tan^{-1}\exp\left[\pm\left(\frac{x-\int_0^t U dt' - X}{(1-U^2)^{0.5}}\right)\right], \tag{A1}$$

where *X* is the initial position of the center of the soliton. In the case of a small perturbation the velocity and position of soliton are described by the formulas ((4.1a) and (4.1b) from (MS)):

$$U_{,t} = \mp \frac{1-U^2}{4} \int_{-\infty}^{\infty} f\,\text{sech}(\theta)\,dx, \tag{A2a}$$

$$X_{,t} = -\frac{U(1-U^2)^{0.5}}{4} \int_{-\infty}^{\infty} f\theta\,\text{sech}(\theta)\,dx. \tag{A2b}$$

where $\theta = \pm \frac{x-\int_0^t U dt' - X}{(1-U^2)^{0.5}}$. (Subscript notations like (,*t*) mean derivation by the respective variable (*t* in this case)).

The perturbed movement of solitons is accompanied by the emission of phonons. Here we are interesting in radiation generated by a single soliton as it responds to the perturbation (single-fluxon case in (MS)). In this case the phonon solution $\vec{W} \equiv \begin{Bmatrix} u(x,t) \\ u_t(x,t) \end{Bmatrix}$ may be expressed through the Green function $G_c$ (see section VI.2 from (MS) formula (6.13)) as:

$$\vec{W} = \int_0^t \int_{-\infty}^{\infty} G_c(x,t|x',t') \vec{F}(x',t') dx' dt', \tag{A3}$$

where

$$G_c(x,t|x',t') = \frac{1}{4\pi i} \int_{-\infty}^{\infty} d\lambda \begin{pmatrix} g_{11} & g_{12} \\ g_{21} & g_{22} \end{pmatrix} \frac{\exp[-i\phi(x-x')-i\varphi(t-t')]}{\lambda(\zeta^2-\lambda^2)^2},$$

$\phi = 2\lambda - 1/(8\lambda)$, $\varphi = 2\lambda + 1/(8\lambda)$,

$\zeta$ relates to $U$ by $U = (16\zeta^2 + 1)/(16\zeta^2 - 1)$,

$g_{11} \equiv (\zeta^2 + \lambda^2 + 2\zeta\lambda\tanh(\tilde{x}))[i\varphi(\zeta^2 + \lambda^2 - 2\zeta\lambda\tanh(\tilde{x}')) - 2i\zeta\lambda\varphi\operatorname{sech}^2(\tilde{x})]$,

$g_{12} \equiv -(\zeta^2 + \lambda^2 + 2\zeta\lambda\tanh(\tilde{x}))[(\zeta^2 + \lambda^2 - 2\zeta\lambda\tanh(\tilde{x}'))]$, ($g_{21}$ and $g_{22}$ will not be used in further calculation so their definitions are omitted),

$\tilde{x} = \frac{x - \int_0^t U dt' - X}{(1-U^2)^{0.5}}$, $\tilde{x}' = \frac{x' - \int_0^{t'} U dt'' - X}{(1-U^2)^{0.5}}$, and

$$\vec{F}(x,t) = \begin{Bmatrix} -u_{0,U} U_{,t} + u_{0,x} X_{,t} \\ f - u_{0,t,U} U_{,t} + u_{0,t,x} X_{,t} \end{Bmatrix},$$

where $u_0$, $U$ and $X$ are defined by the equations (A1) and (A2).

Note that without loss of generality the initial velocity of the soliton may be taken equal to zero ($U=0$), the same is true for the initial position of the soliton ($X=0$). Supposing that $f = a_{init}\delta(x)\delta(t)$ and considering the first term of the vector $\vec{W}$ we find from the equation (A3) after integration:

$$u(x,t) = u_1^+ + u_1^- + u_2^+ + u_2^-, \tag{A4}$$

where

$u_1^+ = \frac{-1}{4\pi i} \int_{-\infty}^{\infty} d\phi \, ((\psi^+)^2 - 1 + 2i\psi^+ \tanh(x))((\psi^+)^2 - 1) \frac{\exp[-i\phi x - i\varphi t]}{\varphi(1+(\psi^+)^2)^2}$,

$u_1^- = \frac{1}{4\pi i} \int_{-\infty}^{\infty} d\phi \, ((\psi^-)^2 - 1 + 2i\psi^- \tanh(x))((\psi^-)^2 - 1) \frac{\exp[-i\phi x + i\varphi t]}{\varphi(1+(\psi^-)^2)^2}$,

$u_2^+ = \frac{1}{8i} \int_{-\infty}^{\infty} d\phi \, \operatorname{sech}(\frac{\phi\pi}{2})((\psi^+)^2 - 1 + 2i\psi^+ \tanh(x))((\psi^+)^2 - 1 + \phi\psi^+) \frac{\exp[-i\phi x - i\varphi t]}{\varphi(1+(\psi^+)^2)^2}$,

$u_2^- = \frac{-1}{8i} \int_{-\infty}^{\infty} d\phi \, \operatorname{sech}(\frac{\phi\pi}{2})((\psi^-)^2 - 1 + 2i\psi^- \tanh(x))((\psi^-)^2 - 1 + \phi\psi^-) \frac{\exp[-i\phi x + i\varphi t]}{\varphi(1+(\psi^-)^2)^2}$,

here $\psi^+ = \phi + \varphi$, $\psi^- = \phi - \varphi$ and $\varphi^2 = \phi^2 + 1$;

Now supposing that $f = a_{init}\eta(x)\delta(t)$ we find from equation (A3) after integration (to distinguish $u(x,t)$ from the previous source (formula (A4) we will use $u_h(x,t)$ for the relative shift of frictional surfaces):

$$u_h(x,t) = u_{h1}^+ + u_{h1}^- + u_{h2}^+ + u_{h2}^-, \qquad (A5)$$

where

$$u_{h1}^+ = \frac{-1}{4\pi}\int_{-\infty}^{\infty} dk(-1+\psi_+^2+2i\psi_+\tanh(x))(-1+\psi_+^2)\frac{(\exp[-i\omega t]-1)\exp[-ikx]}{\omega^2(1+\psi_+^2)^2},$$

$$u_{h1}^- = \frac{-1}{4\pi}\int_{-\infty}^{\infty} dk(-1+\psi_-^2+2i\psi_-\tanh(x))(-1+\psi_-^2)\frac{(\exp[i\omega t]-1)\exp[-ikx]}{\omega^2(1+\psi_-^2)^2},$$

$$u_{h2}^+ = \frac{1}{8}\int_{-\infty}^{\infty} dk(-1+\psi_+^2+2i\psi_+\tanh(x))(-1+\psi_+^2+2\psi_+ k)\text{sech}[\frac{k\pi}{2}]\frac{(\exp[-i\omega t]-1)\exp[-ikx]}{\omega^2(1+\psi_+^2)^2},$$

$$u_{h2}^- = \frac{1}{8}\int_{-\infty}^{\infty} dk(-1+\psi_-^2+2i\psi_-\tanh(x))(-1+\psi_-^2+2\psi_- k)\text{sech}[\frac{k\pi}{2}]\frac{(\exp[i\omega t]-1)\exp[-ikx]}{\omega^2(1+\psi_-^2)^2}.$$